\documentclass{emulateapj}
\usepackage{float}
\usepackage{amsmath}
\usepackage{epsfig,floatflt}
\usepackage{subfigure}
\usepackage{color}
\newcommand{\be}{\begin{equation}}
\newcommand{\ee}{\end{equation}}
\newcommand{\ba}{\begin{eqnarray}}
\newcommand{\ea}{\end{eqnarray}}
\newcommand{\brr}{\begin{array}}

\newcommand{\err}{\end{array}}
\newcommand{\bc}{\begin{center}}
\newcommand{\ec}{\end{center}}

\begin{document}

\title{Gamma-ray emission from dark matter wakes of recoiled black holes}

\author{Roya Mohayaee\altaffilmark{1}, Jacques Colin\altaffilmark{2}, Joseph Silk\altaffilmark{3}
   }

\altaffiltext{1}{CNRS, Institut d'Astrophysique de Paris (IAP) UMR 7095, UPMC, 
98 bis boulevard Arago, France}

\altaffiltext{2}{Universit\'e Nice Sophia Antipolis, CNRS, Observatoire de la
C\^ote d'Azur, UMR 6202 (Cassiop\'ee), Nice, France}

\altaffiltext{3}{University of Oxford, Astrophysics, Keble Road, Oxford OX1 3RH, U.K.}

\begin{abstract}

A new scenario for the emission of high-energy gamma-rays 
from dark matter annihilation around massive black holes is presented.
A black hole can leave 
its parent halo, by means of gravitational radiation recoil, in a merger event 
or in the asymmetric
collapse of its progenitor star. 
A recoiled black hole which moves on an almost radial 
orbit outside the virial radius of its central halo, in the cold dark matter background, 
reaches its apapsis in a finite time. 
Near or at the apapsis passage, a high-density wake extending over a large radius of influence,
forms around the black hole. Significant
gamma-ray emission can result from the enhancement of neutralino annihilation in
these wakes. At its apapsis passage, a black hole 
produces a flash of high-energy 
gamma-rays whose duration is determined by the mass of the
black hole and the redshift at which it is ejected. The ensemble 
of such black holes in the Hubble volume is shown to 
produce a diffuse high-energy gamma-ray background whose magnitude is compared 
to the diffuse emission from dark matter haloes alone.


\end{abstract}


\maketitle


\section{Introduction}

We study the effect on dark matter (DM) distribution of black holes (BHs) that are ejected
from their parent haloes but remain on bound orbits outside the virial radius. 
Black holes can be recoiled due to anisotropic emission of gravitational 
radiation (Bekenstein 1973, Fitchett 1983, Merritt et al 2004). 
As an ejected black hole (BH) moves in the dark matter-dominated universe, a {\it wake}
forms whose
density is maximum downstream along the symmetry axis. The
wake density and its extent, {\it i.e.} the zone of influence of the BH, increase 
with decreasing velocity of the BH and velocity dispersion of the DM environment. 
A black hole on bound orbit round its parent halo comes to rest at apapsis of 
its highly radial orbit in a finite time. 
In a cold dark matter background with a velocity dispersion of the
order of cm/s, a BH near its apapsis has a large radius of influence.
If DM were to consist of
self-annihilating particles such as neutralinos, significant $\gamma$-rays would be
emitted from these wakes. 
On its apapsis passage, a BH is shown to produce a 
flash of high energy $\gamma$-rays whose duration depends
on the mass of the BH and the redshift at which it is ejected. 
We also estimate the total isotropic diffuse gamma ray background emitted from
the ensemble of cosmological wakes in the Hubble volume, assuming that only a
small fraction of BHs leave their haloes and compare it to the background from
DM haloes alone.

There are two main assumptions made in this work. First, we assume that 
outside haloes, the cold dark
matter environment is approximately homogeneous: an
assumption which is more valid at high redshifts. 
Second, to evaluate the wake density we assume 
a constant-velocity approximation, whose validity is
questionable and remains to be confirmed by future N-body simulations.

\section{Density of the wake}
\label{sec:density}

The expression for the density
enhancement due to a moving point mass in a thermal environment is 
(Danby \& Camm 1957, Griest 1988)
\be
{\rho\over\bar\rho}\!=\!\int_{u=0}^{\infty}\!\!\!
du\,{u\sqrt{u^2+q^2}\over (2\pi)^{3/2}}\int_{\lambda=0}^\pi\!\!\!\!\!\! {\rm
  sin}\lambda\,d\lambda\int_{\nu=0}^{2\pi}\!\!\!\! d\nu\, e^{-F/2}
\label{eq:density-griest}
\ee
where
$
F=p^2+u^2+2pu
(uZ+q^2{\rm cos}\theta/2-\,Z\sqrt{u^2+q^2}{\rm cos}\lambda)
/
(u^2+q^2/2-u{\rm cos}\lambda\sqrt{u^2+q^2}),
$
$p=V_{\rm BH}/\sigma_{\rm DM}$, $q=2GM_{\rm BH}/r/\sigma_{\rm DM}^2$, 
$Z=\sqrt{u^2+q^2}({\rm cos}\theta{\rm cos}
\lambda-{\rm sin}\theta{\rm sin}\lambda{\rm sin}\nu)$,
$\bar\rho$ is the density of the environment of the BH,
$M_{\rm BH}$ is the mass of the BH moving with velocity $V_{\rm BH}$ and the 
distance $r$ and angle $\theta$ 
are the radial distance measured from the BH
position and the angular position is measured from the symmetry axis with
$\theta=0$ upstream in front of the BH and $\theta=\pi$ downstream on the symmetry axis behind
the BH.

When the velocity dispersion of the medium is very low 
in the limit $V_{\rm BH} >> \sigma_{\rm DM}$, the integral in (\ref{eq:density-griest})
becomes highly oscillatory and difficult to evaluate. The expression for
the density enhancement in this limit (${V_{\rm BH}>>\sigma_{\rm DM}}$) is (see Sweatman \&
Heggie 2004 for details, also Sikivie \& Wick 2002 for a different approach)
\be
{\rho\over\bar\rho}\!=\!
{2\,V_{\rm BH}^2\over \pi\sigma_{\rm DM}^2}
\int_0^\infty\!\!\!\!\int_{\alpha=0}^\pi\!\!\!\!\!\!\!\! n\,dn 
\, d\alpha\,
e^{-V_{\rm BH}^2n^2/\sigma_{\rm DM}^2} {f\over\sqrt{f^2-1}}
\label{eq:density-sweatman}
\ee
where 
$
f=
1+(rV_{\rm BH}^2)/ (2GM_{\rm BH})\times(1+{\rm cos}\theta-2\sqrt{1+{\rm cos}\theta} \,
n\,{\rm cos}\alpha+n^2)\,.
$
the density (\ref{eq:density-sweatman}) along 
the symmetry axis ($\theta=\pi$) attains the maximum value:
$
\rho_{|axis}\approx{\bar\rho(z)\over\sigma_{\rm
DM}(z)}\sqrt{\pi\,G\,M_{\rm BH}/ r}
\label{eq:densityaxis}
$
for $\theta=\pi$  and $\sigma_{\rm DM}^2<<GM_{\rm BH}/r$.
The wake density is independent of the velocity of the
BH, downstream along the symmetry axis
due to the finite velocity dispersion of dark matter.

\section {Density enhancement and radius of influence: the r\^ole of
BH velocity}

\begin{figure}
\includegraphics[width=0.9\columnwidth]{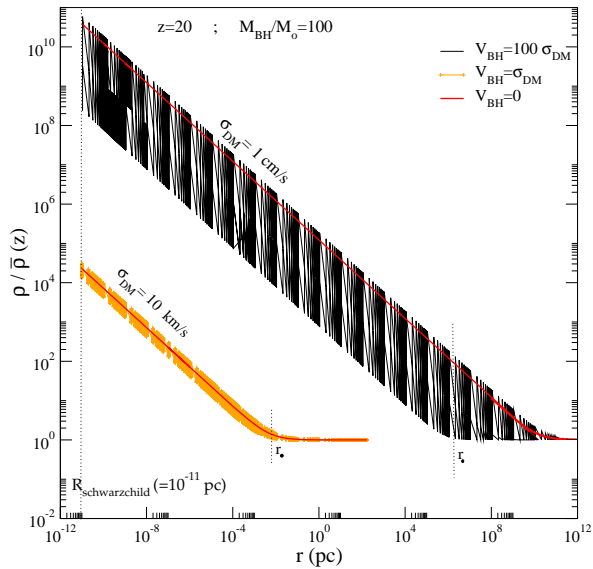}
\caption
{
The change in the wake overdensity (\ref{eq:density-griest}) with 
radial distance $r$ from BH in a hot environment (lower curve) and cold
environment (upper curve). The 
lower and upper envelopes to the black bands
correspond to $\theta=0$ and $\theta=\pi$ respectively.
The solid (red) curves are for stationary BHs (\ref{eq:density-stationary}). 
}
\label{fig:zone of influence}
\end{figure}
\begin{figure}
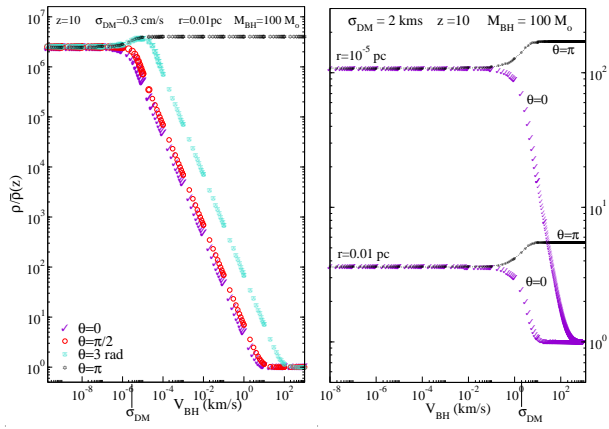

\centering{
\includegraphics[width=0.47\columnwidth]{Density-V-WARM-HOMOGENEOUS-z=10-M=100-r=0o01pc-2}
\includegraphics[width=0.45\columnwidth]{Density-V-WARM+HOT-HALO-z=10-M=100-2}
}
\caption
{
The density (\ref{eq:density-griest}) for $V_{\rm BH}\le \sigma_{\rm DM}$
and (\ref{eq:density-sweatman}) for $V_{\rm BH}>>\sigma_{\rm DM}$ are shown
for a BH moving in a cold ($\sigma_{\rm DM}=0.03 {\rm cm/s}$) [left panel] 
and hot  ($\sigma_{\rm DM}=2 {\rm km/s}$) [right panel] environment. 
}
\label{fig:density-v}
\end{figure}

The zone of influence of BH decreases with increasing velocity and
velocity dispersion of its environment as shown in
Figs.~\ref{fig:zone of influence} and \ref{fig:density-v}. In 
Fig.~\ref{fig:zone of influence}, produced by numerical integration of
equation (\ref{eq:density-griest}), the 
red middle curve for $V_{\rm BH}=0$  shows that 
this approximation provides a good description of the average wake density of
slowly-moving BHs. In the limit as $V\rightarrow 0$, the
density profile of the wake (\ref{eq:density-griest}) reduces to
\be
{\rho\over \bar\rho}= \sqrt{4\over\pi}\sqrt{r_\bullet\over \,r}+
e^{r_\bullet/r}{\rm Erfc}\left(\sqrt{r_\bullet\over r}\right),
\label{eq:density-stationary}
\ee
where ${\rm Erfc}$ is the complementary error function and $r_\bullet$ is the radius of influence
of the BH: $r_\bullet=GM_{\rm BH}/\sigma_{\rm DM}^2$. We emphasis 
that (\ref{eq:density-stationary}) is the limit $V\rightarrow 0$ of
(\ref{eq:density-griest}), and is not a unique density profile for stationary
BHs. Here, we use (\ref{eq:density-stationary}) only as an approximation to
(\ref{eq:density-griest}) for slowly-moving BHs.
Fig.~\ref{fig:density-v}, shows the
dependence of the density enhancement on the BH velocity and DM
velocity dispersion. When the BH is moving fast with respect to the background, a significant
density enhancement only arises in a small zone around the symmetry axis
(downstream) of the BH. The density enhancement also decreases with
increasing velocity dispersion of dark matter environment. 
The highest density enhancement and largest radius of influence are achieved for BHs
moving slowly in a cold background. 

\section{Drag forces on the ejected BH: time taken to reach the apapsis}

The BH remains bound to its central halo if 
it is ejected with a velocity less 
than the escape velocity (measured from the virial radius). 
Because it is ejected from the centre (and also when with a large velocity),
the BH is on almost radial orbit.
The BH initial velocity is set as follows.
We assume that the halo mass is about $2\times\, 10^4$ times 
the mass of the BH (Madau \& Rees 2001). Thus,
for a BH of mass $M_{\rm BH}$, the virial radius of the halo, from which it was
ejected, can be determined
using $M_{\rm halo}={4\pi/3}\Delta_{\rm vir}(z)\bar\rho(z)R_{\rm vir}^3(z)$
and noting that $\Delta_{\rm vir}(z)=(18\pi^2+82x-39x^2)/\Omega(z)$ and
$x=\Omega(z)-1$ and  
 $\Omega(z)=\Omega_m(1+z)^3/[\Omega_m(1+z)^3+\Omega_\Lambda+\Omega_k(1+z)^2]$
(see Bullock et al 2001 for details). Having evaluated the
virial radius, we can then evaluate the escape velocity from the
virial radius of the halo, using $V_{\rm escape}(z,M)=\sqrt{2GM_{\rm halo}/R_{\rm vir}}$. 

\begin{figure}
\includegraphics[width=0.97\columnwidth]{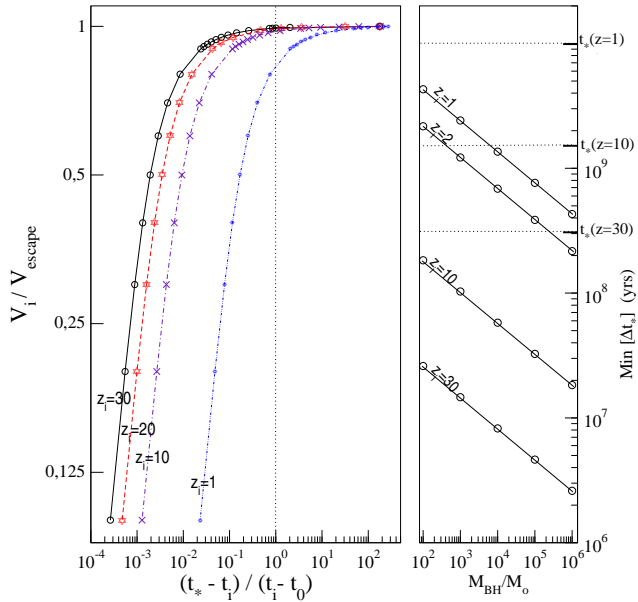}
\caption
{
{\it Left panel:}
The ratio of initial BH velocity at 
the virial radius, $V_i$, to escape
velocity, versus  ratio 
of time of ejection of BH, $t_i$, 
to when it
reaches apapsis, $t_*$, to the time left from the moment 
of ejection  to now ($t_0-t_*$). 
{\it Right panel:} Minimum time the BH spends around 
the apapsis when its
luminosity dominates over that of its parent halo [using expression
(\ref{eq:minimumalpha}) in (\ref{eq:timeatapapsis})].
The total time, $t_*$, taken to reach the apapsis from the virial radius is
also shown by the dotted horizontal lines for different redshifts, for BHs
ejected with near escape velocity.
}
\label{fig:TIME}
\end{figure}

The ejected BH is slowed down by the gravitational pull of its parent halo 
and also by the dynamical friction of 
dark matter background as
\be
{dV_{\rm BH}\over dt}\!=\!-\left[{(2\,E+V_{\rm BH}^2)^2\over 4\,G\,M_{\rm halo}}
+{4\,\pi\,G^2\,
M_{\rm BH}\bar\rho {\rm ln}(\Lambda) \over V_{\rm BH}^2} \right]\!\!\!\!\!
\label{eq:force}
\ee
where $E=-V_i^2/2+G\,M_{\rm halo}/R_{\rm vir}$ is the 
absolute value of the energy
with which a bound BH leaves the virial radius with velocity $V_i$.
Since dynamical friction plays a sub-dominant r\^ole in {\it braking} the BH,
the values of ${\rm ln}(\Lambda)$ and the background density 
$\bar\rho$ marginally affect the value of (\ref{eq:force}) for a BH in its
initial outward journey. 
We have set ${\rm ln}(\Lambda)$ to unity and 
$\bar\rho$ to $\bar\rho_0$ in order to obtain an upper value on the time 
to apapsis. The left panel of Fig.~\ref{fig:TIME}, produced from
numerical integration of equation (\ref{eq:force}) demonstrates
that the time elapsed, $t_*-t_i$,  since the BH was ejected till 
it reaches the apapsis, is far shorter than the 
Hubble time, $t_0$, as long as the BH escapes the virial 
radius with a velocity less than the escape velocity. This figure 
also demonstrates that $t_*-t_i$ is 
independent of the BH mass, because higher mass BHs 
need larger velocities to leave their more massive haloes.

\section{Gamma-ray flashes from BHs around apapses} 
\label{sec:Phi1BH}
Cold dark matter if composed of neutralinos would self-annihilate into secondary
products, including energetic photons (e.g. see Bertone, Hooper, Silk
2005 for a recent review).
The {\it absolute  luminosity}, in units of $\gamma {\rm s}^{-1}\,$ 
of a BH of mass $M$ at redshift $z$ is:
\be
L (M,z)\!\!=\!\!\left[{\cal N}_\gamma{\langle\sigma v\rangle\over 2m_\chi^2}\,\right]\!\!\!
\left[4\,\pi\int_{r_s}^{r_\bullet}\!\!\!\!\! r^2 dr  
\left(\rho(M,z,r)\right)^2\,\right]
\label{eq:L}
\ee
where $r_s$ is the Schwarzchild radius, $r_\bullet$ is the radius of influence of the
BH, $m_\chi$ is the neutralino mass ($\sim 100$ Gev/c$^2$), 
$\langle\sigma v\rangle$ is the interaction cross-section [which we fix at  
$2\,\times\,10^{-26}$ cm$^3$/s] and ${\cal N}_\gamma$ is 
the number of photons produced per annihilation.
Note that the integral
(\ref{eq:L}) is independent of angle $\theta$ for stationary BHs.

We had found that the wake density of a slowly moving BH is well-approximated 
by the wake density of a stationary BH. 
By inserting (\ref{eq:density-stationary}) [keeping only the first
term] in (\ref{eq:L}) we obtain the analytic
expression (in units of $\gamma\, s^{-1}$)
\be
L_{\rm BH}=1.4\,\times\,10^{25}\,(1+z)^4\left({M_{\rm BH}\over
  M_\odot}\right)\,R_{\rm cutoff}^2
\label{eq:LBH}
\ee
for the absolute luminosity of a BH, where $R_{\rm cutoff}$ is in unit of parsecs. In (\ref{eq:LBH})
$R_{\rm cutoff}$ is the radius within which the luminosity of the BH is
evaluated. 

Next, the BH luminosity is compared to the absolute luminosity of its central dark matter halo
 of mass $2\times 10^4\, M_{\rm BH}$, assuming it has a NFW density profile 
(Navarro, Frenk \& White 1997)
\be
{\rho\over \bar\rho_0}=(1+z)^3{\Omega_m\over \Omega(z)}{\delta_c\over
  [(c\,r/R_{\rm vir})(1+(c\,r/R_{\rm vir}))^2]}
\label{eq:NFW}
\ee
where 
$\Omega(z)=\Omega_m(1+z)^3/[\Omega_m(1+z)^3+\Omega_\Lambda+\Omega_k(1+z)^2]$,
$\delta _c=(\Delta_{\rm vir}/3) c^3/({\rm ln}(1+c)-c/(1+c))$,
 and $c$ is 
the concentration parameter, for which we use the fit
$[10/(1+z)] (M_{\rm vir}/M_{\*})^{-0.13}$ (agrees well with
Bullock et al 2001, Hennawi et al 2007) and  
$M_{\rm halo}= M_{\rm vir}$ is the halo mass within the virial radius.
The absolute luminosity  $L$ (\ref{eq:L}), in unit of $\gamma\,s^{-1}$,
of a NFW halo of mass $M_{\rm
halo}$ can then be evaluated  using (\ref{eq:NFW}) in (\ref{eq:L}) [after
setting $M=M_{\rm halo}, r_\bullet=R_{\rm vir}, r_s=0$] 
and fitted by the following functions
\ba
L_{\rm NFW} (z>1)  &=&  
{5.6\times 10^{27}\over (1+z)^{-1/3}}\left({M_{\rm halo}\over
  M_\odot}\right)^{0.7\,(1+z)^{0.075}}\nonumber\\
L_{\rm NFW} (z\le 1) &=& {1.6\times 10^{29}
\over (1+z)^{3}}\left({M_{\rm halo}\over
  M_\odot}\right)^{0.7}\,
\label{eq:LNFW}
\ea

\begin{figure}
\includegraphics[width=0.97\columnwidth]{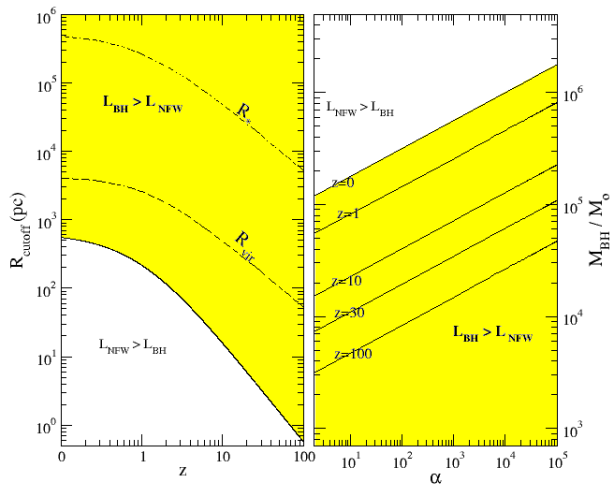}
\caption
{
{\it left panel:}
The minimum radius $R_{\rm cutoff}$ at apapsis required for the BH luminosity to
dominate over that of its parent halo is shown by the lower solid line. The 
minimum virial radius, $R_{\rm vir}$, of the halo (corresponding to ejection of a BH of mass
$100 M_\odot$) and similarly the minimum distance from the virial radius of the apapsis
$R_*$ are also plotted. {\it right panel:} the minimum value of 
the parameter $\alpha$ as a function of redshift and mass of the ejected BH
[expression (\ref{eq:minimumalpha})] are shown.
}
\label{fig:luminosity-field}
\end{figure}

We can find an analytic expression for $R_{\rm
 cutoff}$ by using the dynamical friction formula (\ref{eq:force}) and
assuming that near the apapsis dynamical friction dominates over the
gravitational pull of the halo. This yields 
\be
R_{\rm cutoff}={\sigma_0^4(1+z)\alpha^4\over 16\pi G^2 M_{\rm BH}\bar\rho_0}
\,\,;\,\,
\Delta t_*\!=\!{\alpha^3\sigma_0^3\over 12\pi G^2 M_{\rm BH} \bar\rho_0}
\label{eq:timeatapapsis}
\ee
where $\Delta t_*$ and $R_{\rm cutoff}$ are
defined as the time-duration and distance-parcours during which the velocity
of BH reduces from a fraction $\alpha$ of the background velocity dispersion
to zero (at the apapsis), i.e.  $0<  V_{\rm BH} < \alpha\sigma(z)$ 
where $\sigma(z)=(1+z)\sigma_0$ and $\sigma_0$ is present velocity dispersion
of DM (approximately 0.03 cm/s for neutralinos).
The lower-bound on  $R_{\rm cutoff}$ and $\Delta t_*$ is obtained by
requiring that  $L_{\rm BH}/L_{\rm NFW}>1$ where $L_{\rm BH}$ is given by
(\ref{eq:LBH})
and $L_{\rm NFW}$ is given by (\ref{eq:LNFW}). These lower bounds are 
shown in Fig.~\ref{fig:luminosity-field}
and Fig.~\ref{fig:TIME} respectively. Fig \ref{fig:luminosity-field} shows 
that at high redshifts most
ejected BHs are more luminous than their parent haloes whereas at low
redshifts a much higher $R_{\rm cutoff}$ is required.  
Similarly, we put a lower limit
on the parameters $\alpha$
\be
\alpha >  {7\times 10 ^{4}\over (1+z)^{0.7}}{\left(M_{\rm BH}\over M_\odot\right)}^{1/4}
\label{eq:minimumalpha}
\ee
for $z>1$ and
for $z\le 1$ we can use the very similar expression 
$\alpha > (10 ^{5}/ (1+z)^{9/8})
(M_{\rm BH}/ M_\odot)^{1/4}$ which 
are shown in Fig.~\ref{fig:luminosity-field}. The two panels of this figure
show that less massive BHs at higher redshifts are the most luminous.
\section{Diffuse gamma-ray background }
\label{sec:totalPhi}

In a cosmological scenario, ejected BHs near their apapses pasages, especially 
those at high redshifts, where the merger rate is
higher, can yield an observable diffused background flux. 
The total flux is given by the integral
\be
\Phi=\int_M\int_z{L(M,z)\over 4\pi r(z)^2}\,N(M,z)\,dM\,d{\cal V}(z)
\label{eq:Phi}
\ee
where $M$ can be either the BH mass ($M_{\rm BH}$) or the halo
mass ($M_{\rm halo}$),
$r(z)=R_H\left(1-{1\over \sqrt{1+z}}\right)$ with Hubble radius
$R_H=4000 {\rm Mpc}$, and $N(M,z)$ is the number density of
the BHs [or haloes in the calculation for NFW haloes] 
and the luminosity of a single BH $L(M,z)$ (or the parent NFW haloes at z) is given 
by (\ref{eq:LBH}) [or (\ref{eq:LNFW})] and the volume element is 
$d{\cal V}={\sin}\psi\,r(z)^2 dr(z)  d\psi\,\,d\varphi$.

The physical number density of the BHs is assumed to follow the Press-Schechter
formalism (Press \& Schechter 1974, Bower 1991),
multiplied by the {\it relative} time a BH spends at 
apapsis, {\it i.e.} $(\Delta t_*)/t_0$ where $t_0$ is the age of the Universe. 

The time spent at apapsis is itself a function of 
BH mass as given by (\ref{eq:timeatapapsis}) and shown in the right panel of
Fig.~\ref{fig:TIME} for minimum value of $\alpha$ [obtained by substituting 
(\ref{eq:minimumalpha}) in (\ref{eq:timeatapapsis})].
Thus only the fraction $\Delta t_*/t_0$ of 
the ejected BHs can be considered to be
actually on their apapses passage in a Hubble time.
We can evaluate the
total flux
by performing the integrals in (\ref{eq:Phi}) 
[multiplied by ${\rm Min}[\Delta t_*]/t_0$ for BHs]. For spectral index $n=-1$ 
and $M_*\sim 10^{12}\,M_\odot$ and for the ensemble of BHs at their apapses
passages and their central haloes 
(assuming NFW profiles) we obtain 
$\Phi_{\rm NFW}\sim
10^{-6}  \qquad \gamma {\rm ~~ cm}^{-2}{\rm sr}^{-1}
$
The flux from the BHs is lower than this value (since we have
evaluated our parameters $R_{\rm cutoff}$ and $\alpha$ by requiring $L_{\rm
  NFW}=L_{\rm BH}$) because the BHs only spend a
fraction of time at the apapsis [Min($\Delta t_*$)] which yields
approximately
$\Phi_{\rm NFW}\sim
10^{-14}  \qquad \gamma {\rm ~~ cm}^{-2}{\rm sr}^{-1}
$
The flux would be attenuated due to interaction of photons which however would affect
{\it approximately} equally $\Phi_{\rm BH}$  and $\Phi_{\rm NFW}$. 
We have assumed that only BHs 
produced in $3\sigma$ peaks of the density perturbation can 
undergo effective mergers. 
We have assumed that all ejected BHs orbit their central haloes outside
the virial radius; 
however, were this not the case, we do
not expect any significant overall decrease in the flux which is already 
underestimated by our moderate choices of parameters 
 and also by assuming that there is
only one apapsis passage for a BH. We have also ignored the effect of multiple 
density-enhancement for a BH which is on its inward journey through an already
high-density wake.

In conclusion,
 BHs on bound orbits around haloes can be powerful sources of
  high-energy $\gamma$-rays, both individually as resolved sources and collectively as
 diffuse background.
The results here indicate that
the globular clusters in the outskirt of our halo or field galaxies in our local 
Universe devoid of central BHs can have 
orbiting BHs which during their apapsis passages 
would produce flashes of high-energy $\gamma$-rays, although
this effect is expected to be most significant at high redshifts.
The validity of 
dynamical friction formulae has been very rarely studied 
for radial orbits (Gualandris \& Merritt
  2007). The fact that there is no mass-loss makes BHs a rare case for
  dynamical friction theory. Throughout this work we have
	assumed a homogeneous background and a constant-velocity 
approximation, both of these assumptions are questionable for the problem
 considered here. 
It thus remains for forth-coming numerical
works to check the extent of the validity of the present results.

{\small This work has been done in collaboration with Michel H\'enon.\
We thank J.P. Lasota,
David Weinberg and the anonymous referee for contributions and French ANR (OTARIE) for
travel grants.}


\end{document}